\documentclass[prc,nofootinbib,showkeys]{revtex4}
\usepackage{amsmath,graphicx}

\begin{document}
%\draft

\title{Fate of QCD sum rules or fate of vector meson dominance in a nuclear medium}

\author{Stefan Leupold}

\affiliation{Institut f\"ur Theoretische Physik, Universit\"at
Giessen, Germany}

\begin{abstract}
A current-current correlator with the quantum numbers of the omega meson is studied
in a nuclear medium. Using weighted finite energy sum rules and dispersion
relations for the current-nucleon forward scattering amplitude it is shown that
strict vector meson dominance and QCD sum rules are incompatible with each other.
This implies that at least one of these concepts --- which are both very powerful
in the vacuum --- has to fade in the nuclear environment.
\end{abstract}
           % PACS numbers from http://publish.aps.org/PACS
           % 10. THE PHYSICS OF ELEMENTARY PARTICLES AND FIELDS
           % 12. Specific theories and interaction models; particle systematics
% 12.38.Lg Other nonperturbative calculations (12.38.-t Quantum chromodynamics)
% 14.40.Cs Other mesons with S=C=0, mass < 2.5 GeV (14. Properties of specific particles)
% 21.65.+f Nuclear matter
% 24.85.+p Quarks, gluons, and QCD in nuclei and nuclear processes

\maketitle

\section{Introduction}
\label{sec:intro}

The present work addresses the interrelation of two important concepts of hadron
physics, namely QCD sum rules \cite{shif79} 
and vector meson dominance (VMD) \cite{sakuraiVMD}. Both concepts are extremely
successful in the description of hadrons and their interactions. However,
if we turn from the elementary (vacuum) properties and cross sections of hadrons
to a many-body system (here: nuclear matter), we will demonstrate that the two
concepts are not compatible any more. We will restrict ourselves here to the
omega meson and comment briefly on possible extensions in the end.
It is important to note that both methods
are also frequently used for in-medium calculations ---
concerning QCD sum 
rules cf.~e.g.~\cite{Hatsuda:1992ez,klingl2,Leupold:1998dg,Thomas:2005dc} and 
references therein, concerning VMD cf.~e.g.~\cite{Gale:1990pn,klingl2} and 
references therein. Our work shows that at least
one of the two methods must be modified (at least for nuclear matter calculations
and at least for the omega meson). In fact, recently the incompatibility of the two 
concepts has been observed numerically in \cite{Steinmueller:2005}. Here we study
that further by presenting an analytical derivation. 

In the next section we will
discuss the QCD sum rule method for our case of interest. In section \ref{sec:vmd}
we introduce the concept of VMD and combine it with the previously derived sum rule
formula. We will end up with an inconsistent relation. This proves that (at least) 
one of the input assumptions --- sum rules or VMD --- must be wrong for the studied
case, omega mesons in nuclear matter. Finally we discuss this issue further in
section \ref{sec:disc}.

\section{QCD sum rules}
\label{sec:cond}

In this work we study the properties of a vector-isoscalar current 
\begin{equation}
  \label{eq:vecisoscal}
j_\mu := \frac12 \left( \bar u \gamma_\mu u + \bar d \gamma_\mu d \right)
\end{equation}
which is at rest with respect to the nuclear medium. From the current-current
correlator
\begin{equation}
  \label{eq:curcur}
\Pi_{\mu\nu}(q) = i \int\!\! d^4\!x \, e^{iqx} 
\langle T j_\mu(x) j_\nu(0) \rangle_{\rm med} 
\end{equation}
we construct the dimensionless quantity
\begin{equation}
  \label{eq:defR}
R(q^2) :=  \frac{\Pi^\mu_\mu(q^2,\vec q\,^2 =0)}{-3 q^2} \,.
\end{equation}
$R$ has a direct physical meaning in the time-like region  $s=q^2 > 0$. 
It is related to the cross section $e^+ e^- \to $ hadrons with isospin 0 via
\cite{klingl2}
\begin{equation}
  \label{eq:crosssec}
\frac{ \sigma^{I=0}(e^+ e^- \to \mbox{hadrons})}{ \sigma(e^+ e^- \to \mu^+ \mu^-) }
= \frac43 \pi {\rm Im}R   \,.
\end{equation}
At least for low energies this cross section is determined by hadronic degrees 
of freedom. For high energies the cross section can be calculated by perturbative QCD.
One supposes that perturbative QCD yields a reliable result for energies above the
so-called continuum threshold \cite{shif79}. One gets
\begin{equation}
  \label{eq:highen}
{\rm Im}R(s) = \frac{1}{8\pi}
\left(1+\frac{\alpha_s}{\pi} \right) \qquad \mbox{for} \quad s > s_0(\rho_N) \,.
\end{equation}
The continuum threshold $s_0$ might depend on the nuclear density $\rho_N$, but
otherwise the high-energy form is assumed to be independent of the medium. This
is in agreement with the general picture that in-medium changes are a (collective) 
long-distance effect. 

As outlined e.g.~in \cite{Hatsuda:1993bv}
in-medium QCD sum rules can be obtained from an off-shell dispersion relation
which integrates over the energy at fixed (here vanishing) three-momentum of the current.
We also restrict ourselves to small densities $\rho_N$ 
by using the linear-density approximation. Effectively this means that the current is at
rest with respect to the nucleon on which it scatters.
The lowest two finite energy sum rules (FESRs) are given 
by \cite{Hatsuda:1995dy,klingl3,Dutt-Mazumder:2000ys}
%(vacuum e.g.~\cite{Chetyrkin:1978ta,Bertlmann:1984ih,Fischer:1986pb,Bertlmann:1987ty,%
%Dominguez:1987nw,Gimenez:1990vg,maltman,Dominguez:1998wy,Davier:1998dz,%
%Hatsuda:1995dy,klingl3,Marco:1999xz,Marco:2001dh,Dominguez:2003dr,Leupold:2004gh}).
%\cite{shif79,Hatsuda:1992ez,klingl2,Dutt-Mazumder:2000ys,Zschocke:2002mp}
\begin{subequations}
  \label{eq:FESR}
\begin{eqnarray}
\label{eq:FESR1}
\frac{1}{\pi} \int\limits^{s_0(\rho_N)}_0 \!\! ds \,
{\rm Im} R(s,\rho_N)  & = &
c_0  s_0(\rho_N) - \frac{9 \rho_N}{4 m_N} \,, \\
\label{eq:FESR2}
\frac{1}{\pi} \int\limits^{s_0(\rho_N)}_0 \!\! ds \, s \,
{\rm Im} R(s,\rho_N)  & = &
\frac12 \, c_0  s_0^2(\rho_N) - c_2  \,.
\end{eqnarray}
\end{subequations}
Here $\rho_N$ denotes the nuclear density and the $c_i$ encode the condensates:
\begin{subequations}
  \label{eq:cond02}
  \begin{eqnarray}
  \label{eq:c0}
  c_0 & = & \frac{1}{8\pi^2}\left(1+\frac{\alpha_s}{\pi} \right) \,, \\
  c_2 & = & m_q \langle \bar q q\rangle_{\rm med} +
  \frac{1}{24} \, \left\langle \frac{\alpha_s}{\pi} G^2 \right\rangle_{\rm med} 
%\nonumber \\ && {}
  + \frac{1}{4} \, m_N a_2 \rho_N \,.
  \label{eq:c2}
  \end{eqnarray}
\end{subequations}
Note that we have followed the common practice to neglect contributions 
from $\alpha_s$ suppressed twist-two operators 
(cf.~e.g.~\cite{Leupold:1998bt} for details). Their inclusion would not change our
lines of reasoning. The quantity $a_2$ is a moment of parton distributions and therefore
well constrained by deep inelastic scattering \cite{Hatsuda:1992ez}.
Note that we have extracted from
$R$ the Landau damping contribution --- last term on the right hand side of
(\ref{eq:FESR1}) \cite{Hatsuda:1992ez}.\footnote{Sign and size of the Landau damping
contribution are discussed in some detail in \cite{Steinmueller:2005}.}

Using the linear-density approximation we get for quark and gluon condensate 
\cite{Hatsuda:1992ez}:
\begin{eqnarray}
m_q \langle \bar q q \rangle_{\rm med} & = & m_q \langle \bar q q \rangle_{\rm vac} 
+ m_q \langle N \vert \bar q q \vert N \rangle \rho_N
%\nonumber \\ 
%& = & 
= -\frac12 F_\pi^2 M_\pi^2 + \frac12 \, \sigma_N \rho_N  \,,
  \label{eq:scal2q}
\end{eqnarray}
and
\begin{equation}
  \label{eq:gluoncond}
\left\langle \frac{\alpha_s}{\pi} G^2 \right\rangle_{\rm med}
= \left\langle \frac{\alpha_s}{\pi} G^2 \right\rangle_{\rm vac}
- \frac{8}{11 - \frac{2}{3}N_f } m_N^{(0)} \rho_N \,.
\end{equation}

Let us briefly discuss the essence of the sum rules (\ref{eq:FESR}): 
The basic idea \cite{shif79} is that the hadronic low-energy spectral information 
(which enters the left hand side) is connected to QCD information about the 
non-perturbative vacuum, or in other words: to perturbative QCD improved by condensates
via an operator product expansion. In short: High-energy information 
(about quarks and gluons) is connected to low-energy information (about hadrons).
As already discussed in \cite{shif79}, it is {\em not} always possible to connect
these informations. At least, one has to make sure that the hadronic side of a sum
rule is indeed given/dominated by the low-energy information and that the
condensate side is dominated by the high-energy information encoded in the known 
condensates of low dimensionality. 
We will come back to that point below.

Besides the condensates which are (more or less) well known and the quantity Im$R$ we are
interested in, we also have the continuum threshold
$s_0$ and its density dependence. It introduces an additional parameter which we are not 
primarily interested in. In addition, the sum rules (\ref{eq:FESR}) mix vacuum and
in-medium information while we are only interested in the latter. At least, we want to
make sure that uncertainties in the vacuum description do not influence our conclusions for
the in-medium changes. In the following, we will show that one can get better sum rules
which solve some of the mentioned problems.

To be more sensitive to the in-medium modifications we differentiate the 
FESRs (\ref{eq:FESR}) with respect to the density. Since we work in
the linear-density approximation we should set $\rho_N$ to zero after the 
differentiation. We get
\begin{subequations}
\label{eq:FESRdens}
\begin{eqnarray}
%\lefteqn{
\frac{1}{\pi} \int\limits_0^{s_0} \!\! ds \, 
\left. \frac{\partial}{\partial \rho_N} {\rm Im}R(s,\rho_N) 
\right\vert_{\rho_N =0} + \tilde c
%} \nonumber \\ && \hspace*{8em}
& = &  c_0 \left. \frac{d s_0}{d \rho_N} \right\vert_{\rho_N =0}
- \frac{9}{4 m_N} \,,  \\[1em]
%\lefteqn{
\frac{1}{\pi} \int\limits_0^{s_0} \!\! ds \, s \, 
\left. \frac{\partial}{\partial \rho_N} {\rm Im}R(s,\rho_N) 
\right\vert_{\rho_N =0} 
+ s_0 \tilde c
%}   \nonumber \\ &&  \hspace*{8em}
& = & c_0 s_0 \left. \frac{d s_0}{d \rho_N} \right\vert_{\rho_N =0} 
- \frac{d c_2}{d \rho_N}  \,. 
%\phantom{mm}
\end{eqnarray}
\end{subequations}
with $s_0 = s_0(\rho_N = 0)$ and
\begin{eqnarray}
  \label{eq:defirr}
\tilde c = \left. 
\frac1\pi \, {\rm Im}R(s_0,\rho_N) \, \frac{d s_0}{d \rho_N}
\right\vert_{\rho_N =0} \,.
\end{eqnarray}

The FESRs (\ref{eq:FESR}) --- and also (\ref{eq:FESRdens}) ---
show an unpleasant feature: The region around the continuum threshold can sizably
contribute to the respective integrals. On the other hand, we recall that the
sum rules are derived under the assumption that at the continuum threshold
perturbative QCD immediately sets in and gives a reliable description. This is
expressed in (\ref{eq:highen}). Clearly this is an over-simplifying assumption.
Therefore, one would prefer sum rules
which are only sensitive to the low-energy region one is interested in and insensitive
to the region around the threshold. This criterion is not met by the FESRs, but e.g.~by
the Borel sum rules
already used in the seminal QCD sum rule papers \cite{shif79}. However, also
Borel sum rules have their shortcomings as compared to FESRs:
In principle, condensates of arbitrary high dimensions enter the Borel sum rules
while only condensates of a specific dimension enter a particular FESR
(e.g.~only dimension-four condensates collected in $c_2$ enter the 
sum rule (\ref{eq:FESR2})). In addition, Borel sum rules introduce additional
parameters (the Borel window) in which one is primarily not interested in. A method which
combines the advantages of FESR and Borel sum rules are weighted FESRs. 
For our case of interest we obtain the following
weighted finite energy sum rule (WFESR) by a linear combination of the 
equations (\ref{eq:FESRdens}):
\begin{eqnarray}
%\lefteqn{
\frac{1}{\pi} \int\limits_0^{s_0} \!\! ds \, (s_0-s)
\left. \frac{\partial}{\partial \rho_N} {\rm Im}R(s,\rho_N)
\right\vert_{\rho_N =0} 
%} \nonumber \\ && \hspace*{8em}
=  - \frac{9s_0}{4 m_N} + \frac{d c_2}{d \rho_N}  \,.
    \label{eq:wFESR1}
\end{eqnarray}
This is the sum rule which we will use from now on.

As already noted, the standard FESRs are rather sensitive to
the modeling of the transition region from the hadronic part 
to the continuum (see also e.g.~\cite{Leupold:2001hj} and references therein). This
is different for (\ref{eq:wFESR1}), since the transition region is suppressed by 
$(s_0-s)$. Therefore (\ref{eq:wFESR1}) is
more reliable as it is insensitive to details of the
threshold modeling at $s_0$. This is a frequently discussed line of reasoning which is
not special for in-medium sum rules 
(cf.~e.g.~\cite{maltman} and references therein).\footnote{Indeed, a very 
interesting and successful application of WFESRs and a failure of standard
FESRs are documented in 
\cite{Dominguez:2003dr} for Weinberg type sum rules
in the vacuum.}
The in-medium WFESR (\ref{eq:wFESR1}), however, has an
additional feature: It is independent of the in-medium change of the threshold parameter,
i.e.~independent of $\frac{d s_0}{d \rho_N}$.  
Note that the vacuum threshold $s_0$
appears in (\ref{eq:wFESR1}). This, however, can be fixed by
an independent vacuum sum rule analysis which is free of all in-medium uncertainties.
If one takes the arithmetic
average of the squared masses of the omega and of its first 
excitation \cite{pdg04} one gets $s_0 \approx 1.3\,$GeV$^2$.
We note already here that for our
qualitative arguments we do not need the numerical value.

We would like to stress again that the sum rule (\ref{eq:wFESR1}) 
constitutes a big step forward in the sum rule analysis of in-medium properties:
First, we are directly sensitive to in-medium changes in contrast to
traditional analyses \cite{Hatsuda:1992ez,klingl2,Leupold:1998dg,Zschocke:2002mp}
which study vacuum plus medium contributions. Second, we have got 
rid of all additional parameters like the in-medium continuum threshold and
the Borel masses 
(cf.~also the discussion in \cite{Hatsuda:1995dy,Mallik:2001gv}). Third, we still share
with Borel type sum rules the feature that we are less sensitive to the modeling of the
continuum threshold.

\section{Vector meson dominance}
\label{sec:vmd}

Strict VMD implies the identification \cite{sakuraiVMD,Gale:1990pn}
\begin{equation}
  \label{eq:strvmd}
  j_\mu = - \frac{M_\omega^2}{g} \, \omega_\mu
\end{equation}
with the mass of the omega $M_\omega \approx 782\,$MeV \cite{pdg04}, 
the universal vector meson coupling $g\approx 6$ \cite{Gale:1990pn}
and the omega meson field $\omega_\mu$. In other words, all hadronic interaction of 
the current (\ref{eq:vecisoscal})
with hadrons is mediated by the omega meson. One gets
\begin{eqnarray}
  \label{eq:vmdR1}
{\rm Im}R(s) & = & -\frac{M_\omega^4}{g^2 s} \, 
{\rm Im} \frac{1}{s-M_\omega^2-\Pi_\omega(s)}
\end{eqnarray}
with the self energy $\Pi_\omega$ 
which in general splits into a vacuum and an in-medium part. 
Let us stress again, that according to strict VMD all in-medium modifications
influence the propagator of the omega meson (via the self energy), 
but not the vertex between the current
(\ref{eq:vecisoscal}) and the omega. In linear-density approximation we get
\begin{eqnarray}
  \label{eq:vmdR}
{\rm Im}R(s,\rho_N) = -\frac{M_\omega^4}{g^2 s} \, 
{\rm Im} \frac{1}{s-M_\omega^2-\rho_N T(s)}
\end{eqnarray}
and therefore
\begin{eqnarray}
%\lefteqn{ 
\left. \frac{\partial}{\partial \rho_N} {\rm Im}R(s,\rho_N)
\right\vert_{\rho_N =0} 
%} \nonumber \\ &&
= \frac{M_\omega^4}{g^2 s} \, {\rm Im} \left[ \left(
\frac{d}{ds} \frac{1}{s-M_\omega^2 + i\epsilon} \right) T(s) \right]
  \label{eq:rdiff}
\end{eqnarray}
with the omega-nucleon forward scattering amplitude $T(s)$ for an in general off-shell
omega with invariant mass squared $s$ which is at rest with respect to the nucleon.
Note that we have neglected the small vacuum self energy of the omega meson. We will
come back to this approximation below.

Using (\ref{eq:rdiff}) we obtain for the left hand side of the 
sum rule (\ref{eq:wFESR1}) after some algebra:
\begin{eqnarray}
  \label{eq:lhs}
%\lefteqn{
\frac{1}{\pi} \int\limits_0^{s_0} \!\! ds \, (s_0-s)
\left. \frac{\partial}{\partial \rho_N} {\rm Im}R(s,\rho_N)
\right\vert_{\rho_N =0} 
%} \\ && 
&= & \frac{M_\omega^4}{g^2 \pi} \left\{
\int\limits_0^{s_0} \!\! ds \, \left[
\frac{s_0}{s^2} \, {\rm Re}\frac{1}{s-M_\omega^2 + i\epsilon} \, {\rm Im} T(s)
%\right. \right. \nonumber \\ && \left. \hspace{7em}
- \left( \frac{s_0}{s} -1 \right) {\rm Re}\frac{1}{s-M_\omega^2 + i\epsilon} \, 
{\rm Im} T'(s) \right] 
\right. \nonumber \\ && \left. \hspace{3em}
- \frac{s_0}{M_\omega^4} \, \pi {\rm Re}T(M_\omega^2)
+ \left( \frac{s_0}{M_\omega^2} -1 \right) \pi {\rm Re}T'(M_\omega^2) \right\} \,.
%\nonumber 
\end{eqnarray}
It is important to note that the integrals which appear in the last expression look 
like dispersive integrals. Indeed, we will use in the following dispersion relations
to calculate Re$T$ and bring the non-integral terms on the right hand side of 
(\ref{eq:lhs}) in a form similar to the integral terms. There
is, however, one notable difference: Dispersive integrals cover the whole energy
range, i.e.~they are not restricted by $s_0$.

Next we relate the real part of the scattering amplitude to its imaginary part
using a one time subtracted dispersion relation, i.e.~\cite{klingl2}
\begin{eqnarray}
{\rm Re}T'(s) & = & - \frac1\pi {\cal P}\int\limits_0^\infty \!\! ds' \,
\frac{{\rm Im}T'(s')}{s-s'} 
%\nonumber \\ & = & 
= - \frac1\pi \int\limits_0^\infty \!\! ds' \, {\rm Im}T'(s') \, 
{\rm Re}\frac{1}{s-s'-i\epsilon}
  \label{eq:dispder}
\end{eqnarray}
and therefore
\begin{eqnarray}
  \label{eq:dispsub}
{\rm Re}T(s) = {\rm Re}T(0) - \frac{s}{\pi} \int\limits_0^\infty \!\! ds' \, 
\frac{{\rm Im}T(s')}{s'}
\, {\rm Re}\frac{1}{s-s'-i\epsilon}  \,. \phantom{m}
\end{eqnarray}

The subtraction constant appearing in (\ref{eq:dispsub}) is the omega-nucleon
forward scattering amplitude at the photon point. VMD relates this quantity
to the isospin-0 part of the photon-nucleon scattering amplitude for vanishing photon
momenta. The latter is just Thomson scattering and therefore we get
\begin{eqnarray}
  \label{eq:thom}
{\rm Re}T(0) = \frac{9 g^2}{4 m_N}  \,.
\end{eqnarray}
Now we use (\ref{eq:lhs}), (\ref{eq:dispder}) and (\ref{eq:dispsub}) 
to evaluate the left hand side of the WFESR (\ref{eq:wFESR1}).
We get after some lengthy but straightforward algebra
\begin{eqnarray}
  \label{eq:nfin}
-\frac{s_0}{g^2} \, {\rm Re}T(0) +
\frac1\pi \frac{M_\omega^4}{g^2} \, X = - \frac{9s_0}{4 m_N} + \frac{d c_2}{d \rho_N}
\end{eqnarray}
with 
\begin{eqnarray}
X & = & \frac{1}{M_\omega^2} \, {\rm Im}T(s_0) 
- \frac{s_0}{M_\omega^2} \int\limits_{s_0}^\infty \!\! ds \, 
\frac{{\rm Im}T(s)}{s}
\, {\rm Re}\frac{1}{s-M_\omega^2+i\epsilon}  
%\nonumber \\ && 
+ \left( \frac{s_0}{M_\omega^2}-1 \right) \int\limits_{s_0}^\infty \!\! ds \,
{\rm Im}T'(s) \, {\rm Re}\frac{1}{s-M_\omega^2+i\epsilon} \,.
  \label{eq:defX}
\end{eqnarray}
Using finally (\ref{eq:thom}) we get
\begin{eqnarray}
  \label{eq:fin}
\frac1\pi \frac{M_\omega^4}{g^2} \, X = \frac{d c_2}{d \rho_N}  \,.
\end{eqnarray}
It is important to note that $X$ contains purely high-energy information, i.e.~the
imaginary part of the scattering amplitude {\em above} the continuum threshold.
Concerning the sum rule philosophy this is a disastrous result: The condensates
on the right hand side --- to be more precise: their in-medium
changes --- do not constrain the low-energy information as they should, but rather
the high-energy information. In contrast, the basic idea of the sum rule method is
to make contact between the condensates and the hadronic low-energy information.
As already discussed above, the high-energy part (and especially the part around the
continuum threshold) is much less accurately treated.
In turn this means that a sum rule becomes unreliable, if it is not the low-energy
information which is connected to the condensates, but part of the high-energy
information. The sum rule method breaks down. 

In the spirit of the sum rule method one might even take a somewhat more simplified
point of view to demonstrate what is wrong with (\ref{eq:fin}): 
We recall the proposition that the high-energy
part of Im$R$ is unchanged by the medium, as expressed in (\ref{eq:highen}).
Consequently, it should not matter much whether one stops the dispersion integral
(\ref{eq:dispsub}) at $s_0$ or at infinity. Stopping the dispersion at $s_0$, however, is
equivalent to completely neglecting $X$. Hence, we would deduce from (\ref{eq:fin}) that
the in-medium changes of the condensates vanish. On the other hand, plugging in 
reasonable numbers \cite{Hatsuda:1992ez} in (\ref{eq:c2}), (\ref{eq:scal2q}), 
(\ref{eq:gluoncond}) one finds that $dc_2/d\rho_N$ does not vanish.
This demonstrates even more clearly the breakdown of the sum rule method.

\section{Further discussion}
\label{sec:disc}

We have found that the sum rule method becomes unreliable for the in-medium
part of the correlator. 
Note, however, that this result is caused by VMD. Without VMD there would be terms
remaining at the left hand side of the sum rules which contain low-energy information.
This could easily be checked e.g.~with the extension used in \cite{Friman:1997tc}.
We will not go through this exercise here.
Hence, the main result of our purely analytical calculations is the following: 
If strict VMD still works well in a nuclear medium, then the sum rule method does
not work any more. In turn, this means, that if the sum rule method works
in a nuclear medium, then strict VMD has to fade. We would like to stress that neither
the sum rule method nor VMD is strictly derivable from QCD. Therefore, we will not
speculate in the present paper which of the concepts should be modified. We only report
our finding that they do not fit together for an in-medium situation.
We note in passing that an in-medium fate of VMD has been found in the context of
the hidden local symmetry approach \cite{Harada:2003jx}. 

We should discuss the methods and approximations used to obtain our result:
By replacing (\ref{eq:vmdR1}) by (\ref{eq:vmdR}) we have neglected the vacuum
self energy of the omega meson, i.e.~its width and the corresponding real part. 
Indeed, without that approximation the integrals
in (\ref{eq:lhs}) would not resemble dispersions. This would induce changes on the
left hand side of (\ref{eq:fin}) which scale with the (vacuum) width of the omega
meson. Do they contain the proper low-energy information which is constrained by
condensates on the right hand side of (\ref{eq:fin})? In principle, this could be
a way to reconcile sum rules and VMD. We give two reasons why this cannot be the case:
From a phenomenological point of view, we observe that the width of the omega is very
small --- about 1\% of its mass \cite{pdg04}. 
Such small modifications cannot account for the condensates.
Also from a more formal point of view this would be unsatisfying: One can study
the sum rules as a function of the number of colors $N_c$ and explore the
large-$N_c$ behavior \cite{'tHooft:1974jz,witten}. In general, the condensates
(\ref{eq:cond02}) are $O(N_c)$. This is also true for their in-medium 
changes.\footnote{The change of the gluon condensate is an exception, 
see e.g.~\cite{Leupold:2004gh} for details.} Also the hadronic left hand side
of the sum rules is $O(N_c)$ which can be seen e.g.~from (\ref{eq:vmdR}) by noting
that the omega mass and the scattering amplitude $T$ are $O(N_c^0)$ and the
coupling $g^2=o(1/N_c)$. The vacuum width of the omega, however, is $o(1/N_c)$. 
Therefore, the inclusion of the vacuum width of the omega on the hadronic side
of the sum rule would induce terms which are $O(N_c^0)$ or lower. This does not
match with the condensate side. We conclude that the inclusion of a vacuum self
energy for the omega does not change our lines of reasoning. We note in passing,
that the large-$N_c$ arguments could be immediately taken over for the rho meson.
Neglecting the vacuum width of the rho meson one would also conclude that
for a rho meson in a nuclear medium strict VMD does not fit together with the sum rules.
On the other hand, from a phenomenological point of view it is less satisfying
to neglect the rather large width ($\approx 150\,$MeV \cite{pdg04}) 
of the rho meson. Therefore we have decided to
concentrate on the omega meson in the present work.

One might ask how sensitive our arguments are concerning the choice of the
used type of sum rules (Borel vs.~FESRs etc.). We have already
motivated why we prefer the use of WFESRs. On top of the arguments already given,
we would like to stress that using a WFESR we were able to show the incompatibility of 
sum rules and VMD in a purely {\em analytical} way. On the other hand, the same 
incompatibility can also be found numerically within analyses using other types of sum 
rules --- and other types of vector mesons.
The sum rule practitioners have just not payed much attention to it. As an example
we discuss the Borel sum rule analysis of \cite{Leupold:1998dg} for the rho meson:
The parameter $F$ introduced in equation 12 in \cite{Leupold:1998dg} should
not change if strict VMD holds. From table 2 of \cite{Leupold:1998dg} we find
that $F$ drops sizably in a medium for all parameter choices. We conclude
that our findings are not special to the chosen type of sum rules. But obviously
the proof is most elegant using a WFESR.

Finally, let us discuss in more detail which kind of VMD is actually incompatible
with the in-medium sum rules. The relation (\ref{eq:strvmd}) which holds on the
level of currents and fields can be obtained from a Lagrangian
\begin{equation}
  \label{eq:lagrsVMD}
{\cal L}_{\rm int} = - \frac{e \, M_\omega^2}{3g} \, \omega^\mu A_\mu
+ \omega^\mu j_\mu^{\rm had}
\end{equation}
with the photon field $A_\mu$, the electromagnetic coupling $e$ and the hadronic 
current $j^{\rm had}_\mu$. 
Such a Lagrangian --- albeit frequently used --- is
somewhat unsatisfying as it mixes photon and vector meson fields. In that way, the
field $A_\mu$ gets a mass and the massless ``real'' photon emerges from a linear
combination of $A_\mu$ and the vector meson field. Such complications are avoided
by a Lagrangian \cite{Friman:1997tc}
\begin{equation}
  \label{eq:lagreVMD}
{\cal L}_{\rm int} = -  \frac{e}{6g} \, \omega^{\mu\nu} F_{\mu\nu}
+ \omega^\mu j_\mu^{\rm had1} + e A^\mu j_\mu^{\rm had2}
\end{equation}
with the field strengths $\omega_{\mu\nu}$ and $F_{\mu\nu}$ for vector meson and
photon, respectively. Now the $A_\mu$ field remains massless, but real photons
decouple from the vector mesons. Therefore, a direct coupling of photons to hadrons
is needed. In general, the Lagrangian (\ref{eq:lagreVMD}) leads to a form for
the current-current correlator which is more complicated than the one 
given in (\ref{eq:vmdR1}). In such a case, one can find in-medium interactions which
do not contradict the sum rule setting. 
However, {\em if} there is a relation between the 
two currents which appear in (\ref{eq:lagreVMD}), namely
\begin{equation}
  \label{eq:had12}
g j_\mu^{\rm had1} = j_\mu^{\rm had2}  \,,
\end{equation}
then relation (\ref{eq:vmdR1}) still holds. Since our proof of incompatibility relies
on (\ref{eq:vmdR1}) and not on the strict VMD relation (\ref{eq:strvmd}) we can conclude
that the sum rules are also incompatible with an in-medium interaction generated from
an extended VMD Lagrangian
(\ref{eq:lagreVMD}), if condition (\ref{eq:had12}) holds.

\bibliography{literature}

\begin{thebibliography}{24}
\expandafter\ifx\csname natexlab\endcsname\relax\def\natexlab#1{#1}\fi
\expandafter\ifx\csname bibnamefont\endcsname\relax
  \def\bibnamefont#1{#1}\fi
\expandafter\ifx\csname bibfnamefont\endcsname\relax
  \def\bibfnamefont#1{#1}\fi
\expandafter\ifx\csname citenamefont\endcsname\relax
  \def\citenamefont#1{#1}\fi
\expandafter\ifx\csname url\endcsname\relax
  \def\url#1{\texttt{#1}}\fi
\expandafter\ifx\csname urlprefix\endcsname\relax\def\urlprefix{URL }\fi
\providecommand{\bibinfo}[2]{#2}
\providecommand{\eprint}[2][]{\url{#2}}

\bibitem[{\citenamefont{Shifman et~al.}(1979)\citenamefont{Shifman, Vainshtein,
  and Zakharov}}]{shif79}
\bibinfo{author}{\bibfnamefont{M.~A.} \bibnamefont{Shifman}},
  \bibinfo{author}{\bibfnamefont{A.~I.} \bibnamefont{Vainshtein}},
  \bibnamefont{and} \bibinfo{author}{\bibfnamefont{V.~I.}
  \bibnamefont{Zakharov}}, \bibinfo{journal}{Nucl.~Phys.}
  \textbf{\bibinfo{volume}{B147}}, \bibinfo{pages}{385, 448}
  (\bibinfo{year}{1979}).

\bibitem[{\citenamefont{Sakurai}(1969)}]{sakuraiVMD}
\bibinfo{author}{\bibfnamefont{J.~J.} \bibnamefont{Sakurai}},
  \emph{\bibinfo{title}{Currents and Mesons}} (\bibinfo{publisher}{University
  of Chicago Press}, \bibinfo{address}{Chicago}, \bibinfo{year}{1969}).

\bibitem[{\citenamefont{Klingl et~al.}(1997)\citenamefont{Klingl, Kaiser, and
  Weise}}]{klingl2}
\bibinfo{author}{\bibfnamefont{F.}~\bibnamefont{Klingl}},
  \bibinfo{author}{\bibfnamefont{N.}~\bibnamefont{Kaiser}}, \bibnamefont{and}
  \bibinfo{author}{\bibfnamefont{W.}~\bibnamefont{Weise}},
  \bibinfo{journal}{Nucl. Phys.} \textbf{\bibinfo{volume}{A624}},
  \bibinfo{pages}{527} (\bibinfo{year}{1997}), \eprint{hep-ph/9704398}.

\bibitem[{\citenamefont{Hatsuda and Lee}(1992)}]{Hatsuda:1992ez}
\bibinfo{author}{\bibfnamefont{T.}~\bibnamefont{Hatsuda}} \bibnamefont{and}
  \bibinfo{author}{\bibfnamefont{S.~H.} \bibnamefont{Lee}},
  \bibinfo{journal}{Phys. Rev.} \textbf{\bibinfo{volume}{C46}},
  \bibinfo{pages}{34} (\bibinfo{year}{1992}).

\bibitem[{\citenamefont{Leupold et~al.}(1998)\citenamefont{Leupold, Peters, and
  Mosel}}]{Leupold:1998dg}
\bibinfo{author}{\bibfnamefont{S.}~\bibnamefont{Leupold}},
  \bibinfo{author}{\bibfnamefont{W.}~\bibnamefont{Peters}}, \bibnamefont{and}
  \bibinfo{author}{\bibfnamefont{U.}~\bibnamefont{Mosel}},
  \bibinfo{journal}{Nucl. Phys.} \textbf{\bibinfo{volume}{A628}},
  \bibinfo{pages}{311} (\bibinfo{year}{1998}), \eprint{nucl-th/9708016}.

\bibitem[{\citenamefont{Thomas et~al.}(2005)\citenamefont{Thomas, Zschocke, and
  K\"ampfer}}]{Thomas:2005dc}
\bibinfo{author}{\bibfnamefont{R.}~\bibnamefont{Thomas}},
  \bibinfo{author}{\bibfnamefont{S.}~\bibnamefont{Zschocke}}, \bibnamefont{and}
  \bibinfo{author}{\bibfnamefont{B.}~\bibnamefont{K\"ampfer}},
  \bibinfo{journal}{Phys. Rev. Lett.} \textbf{\bibinfo{volume}{95}},
  \bibinfo{pages}{232301} (\bibinfo{year}{2005}), \eprint{hep-ph/0510156}.

\bibitem[{\citenamefont{Gale and Kapusta}(1991)}]{Gale:1990pn}
\bibinfo{author}{\bibfnamefont{C.}~\bibnamefont{Gale}} \bibnamefont{and}
  \bibinfo{author}{\bibfnamefont{J.~I.} \bibnamefont{Kapusta}},
  \bibinfo{journal}{Nucl. Phys.} \textbf{\bibinfo{volume}{B357}},
  \bibinfo{pages}{65} (\bibinfo{year}{1991}).

\bibitem[{\citenamefont{Steinm{\"u}ller and Leupold}(2006)}]{Steinmueller:2005}
\bibinfo{author}{\bibfnamefont{B.}~\bibnamefont{Steinm{\"u}ller}}
  \bibnamefont{and} \bibinfo{author}{\bibfnamefont{S.}~\bibnamefont{Leupold}}
  (\bibinfo{year}{2006}), \eprint{hep-ph/0604054}.

\bibitem[{\citenamefont{Hatsuda et~al.}(1993)\citenamefont{Hatsuda, Koike, and
  Lee}}]{Hatsuda:1993bv}
\bibinfo{author}{\bibfnamefont{T.}~\bibnamefont{Hatsuda}},
  \bibinfo{author}{\bibfnamefont{Y.}~\bibnamefont{Koike}}, \bibnamefont{and}
  \bibinfo{author}{\bibfnamefont{S.~H.} \bibnamefont{Lee}},
  \bibinfo{journal}{Nucl. Phys.} \textbf{\bibinfo{volume}{B394}},
  \bibinfo{pages}{221} (\bibinfo{year}{1993}).

\bibitem[{\citenamefont{Klingl and Weise}(1999)}]{klingl3}
\bibinfo{author}{\bibfnamefont{F.}~\bibnamefont{Klingl}} \bibnamefont{and}
  \bibinfo{author}{\bibfnamefont{W.}~\bibnamefont{Weise}},
  \bibinfo{journal}{Eur. Phys. J.} \textbf{\bibinfo{volume}{A4}},
  \bibinfo{pages}{225} (\bibinfo{year}{1999}), \eprint{nucl-th/9901058}.

\bibitem[{\citenamefont{Hatsuda et~al.}(1995)\citenamefont{Hatsuda, Lee, and
  Shiomi}}]{Hatsuda:1995dy}
\bibinfo{author}{\bibfnamefont{T.}~\bibnamefont{Hatsuda}},
  \bibinfo{author}{\bibfnamefont{S.~H.} \bibnamefont{Lee}}, \bibnamefont{and}
  \bibinfo{author}{\bibfnamefont{H.}~\bibnamefont{Shiomi}},
  \bibinfo{journal}{Phys. Rev.} \textbf{\bibinfo{volume}{C52}},
  \bibinfo{pages}{3364} (\bibinfo{year}{1995}), \eprint{nucl-th/9505005}.

\bibitem[{\citenamefont{Dutt-Mazumder et~al.}(2001)\citenamefont{Dutt-Mazumder,
  Hofmann, and Pospelov}}]{Dutt-Mazumder:2000ys}
\bibinfo{author}{\bibfnamefont{A.~K.} \bibnamefont{Dutt-Mazumder}},
  \bibinfo{author}{\bibfnamefont{R.}~\bibnamefont{Hofmann}}, \bibnamefont{and}
  \bibinfo{author}{\bibfnamefont{M.}~\bibnamefont{Pospelov}},
  \bibinfo{journal}{Phys. Rev.} \textbf{\bibinfo{volume}{C63}},
  \bibinfo{pages}{015204} (\bibinfo{year}{2001}), \eprint{hep-ph/0005100}.

\bibitem[{\citenamefont{Leupold and Mosel}(1998)}]{Leupold:1998bt}
\bibinfo{author}{\bibfnamefont{S.}~\bibnamefont{Leupold}} \bibnamefont{and}
  \bibinfo{author}{\bibfnamefont{U.}~\bibnamefont{Mosel}},
  \bibinfo{journal}{Phys. Rev.} \textbf{\bibinfo{volume}{C58}},
  \bibinfo{pages}{2939} (\bibinfo{year}{1998}), \eprint{nucl-th/9805024}.

\bibitem[{\citenamefont{Leupold}(2001)}]{Leupold:2001hj}
\bibinfo{author}{\bibfnamefont{S.}~\bibnamefont{Leupold}},
  \bibinfo{journal}{Phys. Rev.} \textbf{\bibinfo{volume}{C64}},
  \bibinfo{pages}{015202} (\bibinfo{year}{2001}), \eprint{nucl-th/0101013}.

\bibitem[{\citenamefont{Maltman}(1998)}]{maltman}
\bibinfo{author}{\bibfnamefont{K.}~\bibnamefont{Maltman}},
  \bibinfo{journal}{Phys. Lett.} \textbf{\bibinfo{volume}{B440}},
  \bibinfo{pages}{367} (\bibinfo{year}{1998}), \eprint{hep-ph/9901239}.

\bibitem[{\citenamefont{Dominguez and Schilcher}(2004)}]{Dominguez:2003dr}
\bibinfo{author}{\bibfnamefont{C.~A.} \bibnamefont{Dominguez}}
  \bibnamefont{and}
  \bibinfo{author}{\bibfnamefont{K.}~\bibnamefont{Schilcher}},
  \bibinfo{journal}{Phys. Lett.} \textbf{\bibinfo{volume}{B581}},
  \bibinfo{pages}{193} (\bibinfo{year}{2004}), \eprint{hep-ph/0309285}.

\bibitem[{\citenamefont{Eidelman et~al.}(2004)}]{pdg04}
\bibinfo{author}{\bibfnamefont{S.}~\bibnamefont{Eidelman}} \bibnamefont{et~al.}
  (\bibinfo{collaboration}{Particle Data Group}), \bibinfo{journal}{Phys.
  Lett.} \textbf{\bibinfo{volume}{B592}}, \bibinfo{pages}{1}
  (\bibinfo{year}{2004}).

\bibitem[{\citenamefont{Zschocke et~al.}(2003)\citenamefont{Zschocke, Pavlenko,
  and K\"ampfer}}]{Zschocke:2002mp}
\bibinfo{author}{\bibfnamefont{S.}~\bibnamefont{Zschocke}},
  \bibinfo{author}{\bibfnamefont{O.~P.} \bibnamefont{Pavlenko}},
  \bibnamefont{and}
  \bibinfo{author}{\bibfnamefont{B.}~\bibnamefont{K\"ampfer}},
  \bibinfo{journal}{Phys. Lett.} \textbf{\bibinfo{volume}{B562}},
  \bibinfo{pages}{57} (\bibinfo{year}{2003}), \eprint{hep-ph/0212201}.

\bibitem[{\citenamefont{Mallik and Nyffeler}(2001)}]{Mallik:2001gv}
\bibinfo{author}{\bibfnamefont{S.}~\bibnamefont{Mallik}} \bibnamefont{and}
  \bibinfo{author}{\bibfnamefont{A.}~\bibnamefont{Nyffeler}},
  \bibinfo{journal}{Phys. Rev.} \textbf{\bibinfo{volume}{C63}},
  \bibinfo{pages}{065204} (\bibinfo{year}{2001}), \eprint{hep-ph/0102062}.

\bibitem[{\citenamefont{Friman and Pirner}(1997)}]{Friman:1997tc}
\bibinfo{author}{\bibfnamefont{B.}~\bibnamefont{Friman}} \bibnamefont{and}
  \bibinfo{author}{\bibfnamefont{H.~J.} \bibnamefont{Pirner}},
  \bibinfo{journal}{Nucl. Phys.} \textbf{\bibinfo{volume}{A617}},
  \bibinfo{pages}{496} (\bibinfo{year}{1997}), \eprint{nucl-th/9701016}.

\bibitem[{\citenamefont{Harada and Yamawaki}(2003)}]{Harada:2003jx}
\bibinfo{author}{\bibfnamefont{M.}~\bibnamefont{Harada}} \bibnamefont{and}
  \bibinfo{author}{\bibfnamefont{K.}~\bibnamefont{Yamawaki}},
  \bibinfo{journal}{Phys. Rept.} \textbf{\bibinfo{volume}{381}},
  \bibinfo{pages}{1} (\bibinfo{year}{2003}), \eprint{hep-ph/0302103}.

\bibitem[{\citenamefont{'t~Hooft}(1974)}]{'tHooft:1974jz}
\bibinfo{author}{\bibfnamefont{G.}~\bibnamefont{'t~Hooft}},
  \bibinfo{journal}{Nucl. Phys.} \textbf{\bibinfo{volume}{B72}},
  \bibinfo{pages}{461} (\bibinfo{year}{1974}).

\bibitem[{\citenamefont{Witten}(1979)}]{witten}
\bibinfo{author}{\bibfnamefont{E.}~\bibnamefont{Witten}},
  \bibinfo{journal}{Nucl. Phys.} \textbf{\bibinfo{volume}{B160}},
  \bibinfo{pages}{57} (\bibinfo{year}{1979}).

\bibitem[{\citenamefont{Leupold and Post}(2005)}]{Leupold:2004gh}
\bibinfo{author}{\bibfnamefont{S.}~\bibnamefont{Leupold}} \bibnamefont{and}
  \bibinfo{author}{\bibfnamefont{M.}~\bibnamefont{Post}},
  \bibinfo{journal}{Nucl. Phys.} \textbf{\bibinfo{volume}{A747}},
  \bibinfo{pages}{425} (\bibinfo{year}{2005}), \eprint{nucl-th/0402048}.

\end{thebibliography}
\bibliographystyle{apsrev}

\end{document}